# Mediating Community-AI Interaction through Situated Explanation: The Case of AI-Led Moderation

YUBO KOU, Pennsylvania State University, USA
XINNING GUI, Pennsylvania State University, USA

Artificial intelligence (AI) has become prevalent in our everyday technologies and impacts both individuals and communities. The explainable AI (XAI) scholarship has explored the philosophical nature of explanation and technical explanations, which are usually driven by experts in lab settings and can be challenging for laypersons to understand. In addition, existing XAI research tends to focus on the individual level. Little is known about how people understand and explain AI-led decisions in the community context. Drawing from XAI and activity theory, a foundational HCI theory, we theorize how explanation is situated in a community's shared values, norms, knowledge, and practices, and how situated explanation mediates community-AI interaction. We then present a case study of AI-led moderation, where community members collectively develop explanations of AI-led decisions, most of which are automated punishments. Lastly, we discuss the implications of this framework at the intersection of CSCW, HCI, and XAI.

## 1 INTRODUCTION

In recent years artificial intelligence (AI) has been integrated into our everyday technologies, impacting not only individual users, but also communities beyond the individual level, offline and online. AI-powered surveillance cameras are deployed in local communities to track people for security reasons [42]. In many online communities AI has already been used to automatically organize and recommend content [91], and remove harmful content such as hate speech and online harassment [37]. The growing presence of AI in communities is of great interest to HCI and CSCW researchers who have sustained interest in understanding the complex relationship between technology and community as well as how technology could contribute to community wellbeing and development. In this paper, we use the term community-AI interaction to highlight how we as HCI and CSCW researchers could draw from our theoretical and empirical understandings of technology and community to examine AI in the context of community.

**102**

Authors' addresses: Yubo Kou (yubokou@psu.edu) and Xinning Gui (xinninggui@psu.edu), College of Information Sciences and Technology, Pennsylvania State University, University Park, PA, USA 16802

*pre-published version

Particularly, AI systems are often known to be a "black box" with internal mechanisms vague to the users [1,3]. Issues of transparency and accountability have arisen, inspiring a growing body of explainable AI (XAI) [72,97], which has been focused on generating explanations to help ordinary users to understand how AI systems work, make informed decisions, and hold AI accountable [19]. However, existing XAI research has focused largely at the individual level and is cognitive science-driven — how an explanation provides sufficient information for an individual user to understand automated decisions. A majority of technical approaches concern how to develop a simpler technical model of AI for individuals to understand [84]. and HCI researchers have proposed frameworks for user-centered XAI, but the focus is still on understanding how individual users obtain and process information about AI systems [99]. Meanwhile, researchers from other disciplines such as social sciences and philosophy have started to explore the philosophical nature and ontological properties of explanation in AI, pointing to the need of understanding the social, interactive, and dialogical context of explanation [97]. In this regard, explanation is always situated in its specific context. Given that AI has already been applied in real offline and online communities, we see an urgent need in examining XAI in the community context, providing both empirical and conceptual insights into how users as a collective make sense of AI systems deployed in their own community.

Activity theory, a foundational HCI theory that stresses how activity connects mind and its external environment [24,49,75], could offer invaluable insights into the role of explanation in community-AI interaction. In this paper, we draw from activity theory (AT) and existing XAI scholarship to analyze the mediational role of explanation in community-AI interaction. In so doing we seek to illustrate a theoretical foundation and a set of language for reconciling the technical design of XAI system and empirically and theoretically informed HCI. An AT analysis of explanation in community-AI interaction also engenders a concern for contextual and cultural-historical circumstances for explaining that are still missing in current XAI scholarship.

We then use the framework to analyze a case of situated explanation in the research domain of online toxicity and AI-led community moderation. We select this particular research domain because community is built upon norms, shared values, beliefs, and practices [8,20,32,81], and the issue of toxicity directly concerns community norms regarding what content/behavior is acceptable or unacceptable. CSCW and HCI researchers have started to pay more attention to how AI is used to automate the moderation process to curb online toxicity in online communities [37,46], entailing that an AI-led moderation system automatically punishes users who are determined to have violated community norms. However, users could be left in the dark not fully understanding the rationales behind their penalties. This dilemma is precisely what XAI research is proposed to address — to provide sound explanations so that users could understand the rationales behind automated decision making that affect them.

Our study site is *League of Legends*, one of the largest games in the world, developed by Riot Games, Inc. based in Santa Monica, California, United States. The game's moderation system is driven by AI [76,92], where AI algorithms determine whether a behavior is toxic, and issue penalties to convicted players ranging from in-game chat ban to permanent account suspension. Players used a dedicated online forum hosted by Riot Games to seek and generate explanations of their automated penalties. We found that, at the highest level, players sought three primary types of explanations. The first type is socially-oriented explanation, where players focused on figuring out the boundary between toxic behavior and acceptable behavior. The second type is system-oriented explanation, where players paid attention to the particular ways the AI system worked to lead to a decision. The third type is action-oriented explanation, where players sought advice on possible actions following receiving an AI-led decision. These explanations served as knowledge construction processes through which players internalize shared knowledge about behavioral standards and reproduce the social system in the community.

Our work contributes empirical and conceptual insights into the intersection between CSCW, HCI, and AI research: First, we draw from HCI's intellectual and methodological strengths to complement existing research on XAI by articulating the relation of activity theory to XAI research; Second, we describe useful terms and theoretical constructs that could bridge XAI scholarship and HCI scholarship; Third, we provide both conceptual and empirical insights into community-AI interaction; And fourth, we highlight specific ways that XAI designers need to consider when deploying AI systems in pre-existing communities.

## 2. RELATED WORK

### 2.1 Transparency and Accountability of Algorithmic Systems

Transparency means "the visibility of decision-making processes, the clarity with which the reasoning behind decisions is communicated, and the ready availability of relevant information about governance and performance in an organization" [64]. Accountability refers to " the allocation and acceptance of responsibility for decisions and actions and the demonstration of whether and how these responsibilities have been met" [64]. Transparency is commonly seen as a way to promote accountability [71]. Transparency and accountability are important values upheld in the HCI community as researchers examine governance and organization issues in various settings such as open software community [17], online game [73], and healthcare [77].

Existing HCI research has investigated extensively the transparency and accountability of algorithmic systems. Kizilcec et al. [53] and Lee et al. [59] experimented whether transparency could increase users' perceived trust and fairness. Some HCI researchers have examined machine learning practitioners' perspectives and practices on designing transparent and accountable algorithmic systems. Veale et al. [94] investigated public sector machine learning practitioners' perceived challenges understanding and incorporating transparency and accountability into their work. Holstein et al. [44] identified the challenges and needs that machine learning practitioners in commercial product teams face in developing fair and accountable systems.

Explanation has been seen as a primary means to help address these two issues. There has been plenty of work on researcher-driven or designer-driven explanations provided at the user interface. For example, Kitchin discussed ways of researching algorithms, such as examining pseudocode/source code, reflexively producing code, reverse engineering, interviewing designers and conducting an ethnography of a coding team, unpacking the full sociotechnical assemblages of algorithms, and examining how algorithms do work in the world [52]. Rader et al. experimented how they could develop explanations that help participants better understand the internal workings of Facebook's News Feed algorithm [78].

There has been a growing interest in understanding how users themselves develop explanations of algorithmic systems. Eslami and colleagues examined Yelp users' perceptions of the opacity and transparency of algorithms [28], and experimented with Facebook users' awareness of the Facebook News Feed curation algorithm [27]. When algorithmic systems are opaque, end-users may develop folk theories to explain how the underlying algorithms operate and the outcomes and consequences of the systems [18,26,79].

Existing research on user-driven explanations has focused mostly on the individual level. That is, users as individuals develop their own explanations of algorithmic systems. However, users do not always perform cognitive tasks like developing explanation as isolated individuals. They also work as collectives like a community. Plenty of HCI and CSCW research has reported how people utilize online platforms to engage in collective sensemaking to develop syntheses of mixed information and opinions related to health conditions [66] or build shared understandings of cryptocurrency [45].

Collectively, transparency and accountability research has started to pay attention to user-driven explanations, but has not paid much attention to how users as a community could develop explanations. However, considerable HCI and CSCW research has already conceptualized users as collectives working together towards certain cognitive tasks. Bridging these two lines of interest, this study aims to explore how the user community could collectively generate explanations of algorithmic systems.

## 2.2 Explaining Artificial Intelligence (AI)

Current approaches in XAI research are primarily technical. The United States' Defense Advanced Research Projects Agency (DARPA)'s XAI program states explicitly that XAI will create a suite of machine learning techniques that "produce more explainable models, while maintaining a high level of learning performance (e.g., prediction accuracy)" and "enable human users to understand, appropriately trust, and effectively manage the emerging generation of artificially intelligent partners" [39]. Although the DARPA program has a separate psychology track, we argue that significant attention should be paid to the complex, fluid, and nuanced social circumstances that have been the center of CSCW research. In addition, XAI researchers are criticized for their tendency to use their own intuitions as to what constitutes a 'good' explanation [69].

Contemporary philosophy approaches explanation in different traditions such as causal realism, constructive empiricism, ordinary language philosophy, cognitive science, and naturalism and scientific realism [68]. Most relevant to this paper are the first two (causal realism and constructive empiricism), because the rest concern mostly the process of explanation, rather than what explanation is. Causal realists adopt a causal view of explanation, while constructive empiricists accept the causal view as a type of explanation, but view explanation more broadly as an answer to a why-question, both of which are relative and contextual [68]. Causal realism aligns with the notion of XAI the best, because AI as designed system has identifiable processes and components that cause certain outcomes for users. Investigating contextual factors as potential causes aligns with causal realism more, but it would be stretching to adopt constructive empiricism in considering explanation as volatile and free from the inner workings of AI systems. Thus, in this study we adopt a causal view of explanation.

David K. Lewis defines explanation as "to explain an event is to provide some information about its causal history" [63]. According to Lewis, explanation is a process that someone in possession of information about causal relations conveys such information to someone else. An example is to use gravity to explain why an apple falls onto the ground, where gravity is the type of information that has the explanatory power in this case. To explain AI is thus to provide information about what circumstances have caused a particular outcome from an AI system. An example provided by Burrell is that Google's Gmail spam filtering system would provide one reason for why a message is labeled as "spam" such as containing certain keywords [9].

However, human explanation is not meant to fully replicate the machine thinking process, given that the latter is far more complex for human comprehension. Rather, it is to impose "a process of human interpretive reasoning onto a mathematical process of statistical optimization" [9]. Therefore, explaining AI always entails the gap between human comprehension and AI complexity.

A nascent body of HCI research has focused on explaining AI. Cheng et al. experimented how developer-supplied explanations at the interface could help non-expert users [15]. Ashktorab et al. demonstrated user preferences of developer-supplied explanations during chatbot breakdowns [4]. Most relatedly, Wang et al. developed a user-centric XAI framework for the purpose of bettering XAI systems from existing technical, philosophical, and empirical discussions of XAI [99]. However, their theorizations are heavily informed by cognitive sciences and focused on explanation as a form of information. They acknowledge as a limitation that their framework did not consider the social aspect of communication such as cooperative conversation, argumentation, and dialog [99].

Reflecting upon the limitations of the cognitive sciences-informed approaches to XAI, a few researchers have started to examine the contextual nature of human explanation [69,72,97]. Particularly, Miller suggested four properties of explanation that remain underacknowledged in the AI community [69]: explanations could be contrastive, selected, and social, and rely on probabilities the way AI does. Therefore, the accurate presentation of associations and causes is unlikely to suffice human needs. Hilton considered causal explanation first and foremost a form of social interaction [43]. The explainer draws on their experiences, memories, and knowledge, to explain something to the explainee. People seek explanations so as to find meaning and manage social interactions [65]. Building on these discussions of explanation in conversation and argumentation and reflecting on the dominant technical approach in current XAI research, Mittelstadt et al. suggested that XAI research should shift to interactive methods for post-hoc interpretability, while facilitating conversations between users, developers, AI systems, and other stakeholders [72].

In sum, the cognitive-science-informed theorizations of XAI view explanation as information and designing XAI systems as providing such information to users. As a result, scholars from other fields have started to reflect upon the situated nature of explanation. That is, how explanation is related to its context beyond individual users' interaction with AI systems. HCI and CSCW researchers have long investigated how context is related to technology use, applying theoretical frameworks such as situated action [89] and activity theory [5]. At the intersection of XAI and HCI research, we thus found an urgent need in bridging these two research areas by looking into how foundational HCI theories that take context into account could be productively used to analyze XAI phenomena and yield meaningful design directions.

## 2.3 Moderation and its Post-Punishment Challenge

Recently researchers have found renewed interest in understanding approaches to and theories of moderation for many contemporary concerns such as the sheer scale of population residing on one single platform (e.g., Facebook, Reddit, and YouTube) [35], new forms of user-generated content [48], novel moderation tools [46], procedural justice [29], and scholarly caution to the murky role platform owners [82].

Moderation is commonly implemented as a cycle composed of several steps: First, a user's behavior is flagged, and a moderation system adjudicate on the flagged behavior. The user, once convicted by the moderation system, receives punishment of various forms ranging from post removal [47] to account suspension [56]. Moderation could take various sociotechnical forms in terms of human-machine configurations, distribution of power, and proceduralization [88]: Manual content moderation relies upon professional and volunteer human labor, which is commonly found with social media platforms outsourcing moderation tasks to contract employees or crowd workers. Automated content moderation involves the use of automated tools to detect, flag, and remove particular content. An example is to automatically remove content based on regular expressions [46]. But regardless of the forms, moderation systems usually leave punished users on their own to deal with punishments, such as figuring out the rationale behind the penalty and ways of behavioral improvement.

Within the extensive moderation literature, only a few exceptions have paid attention to user understandings of and reactions to punishment. For example, Gerrard's study of eating disorder (ED)-related content on Instagram showed how users learned about moderation criteria and consciously circumvented content moderation to access and circulate pro-ED content [34]. Feuston et al. reflected upon how moderation practice marginalize minority groups [31]. JHaver et al. researched the effect of providing explanations about content removal on Reddit [47]. Most relevant to this research is Kou and Nardi's 2014 study of governance in League of Legends [58], which analyzed how the community helped punished players by explaining their violations in light of community norms and platform rules.

The lack of attention to punishment reflects issues of transparency and accountability in moderation systems. For example, newcomers might commit disruptive behavior because they do not understand the particular community norms [30]. In this case, newcomers who have received penalties need opportunities of learning to improve (or simply modify) their behavior in order to be socialized into the online community. Established communities like Wikipedia with rigid moderation structure might discourage newcomers [40] and marginalize voices of minority groups [98].

In sum, AI-led moderation provides a pertinent scenario for understanding how a community interacts with an AI system that makes important decisions about the community's development and wellbeing. The lack of explanations in moderation design, as is usual for many AI systems, echoes the core concerns of XAI research and our motivation for providing an empirical account of how platform users generate situated explanations for automated decisions.

## 3. THEORETICAL FRAMEWORK: ACTIVITY THEORY AND EXPLANATION OF AI

Activity theory (AT), a framework developed by Soviet psychologists Lev Vygotsky [96] and Alexei Leont'ev [62], was introduced to the HCI community in the early 1990s [16], to analyze human interactions with artifacts and environments [7,86]. AT, deemed as one of the post-cognitivist theories (e.g., distributed cognition and actor-network theory) in HCI [51,83], puts heavy emphasis on the cultural-historical conditions of human activity beyond social aspect of human activities.

At the heart of an AT analysis is the emphasis on how activity plays a central role in mediating between humans and their environments, as well as in the development of human mind [50]. In AT, people as *subjects* in an environment use *tools* to construct and instantiate their intentions and desires as objects [50] (p.10). People do not directly act upon their objects, because *tools*, designed and used in context, mediate their actions. Tools carry cultural knowledge and social experience [75]. In the case of explaining AI, AI is embedded as people's external environments, and people rely upon explanation to develop understandings of this complex external environment. Explanation is an activity that could be initiated by either users or developers of AI and could be mediated by heterogeneous tools. Digital tools such as social media allow people to discuss explanations of AI, while discursive and rhetorical tools in interpersonal communication facilitate people to exchange ideas and information.

An activity contains three layers [50] (p.62): an *activity* with a motive that is the object; an activity includes *actions* that are directly at goals; and each action could be further decomposed into *operations* that routine and sometimes unconscious processes. AI could be involved in human activity at all three layers [93]: AI could augment operations to increase their efficiency and effectiveness; automate actions that were previously done by humans; or significantly transform the current systems of motives and participation at the activity level. In this paper, due to space limit we only consider the activity-level involvement where novel applications of AI in areas like moderation engender entirely new activity systems.

Activities like explanation are collective phenomena with regard to both object and form [50] (p.99). The implication is that explanation should be viewed as a social process rather than merely an isolated individual encounter. Through explanation, community members collectively construct knowledge about AI systems. In particular, AT researcher Engestrom developed the "triangular" *activity system* [22,25], described in Figure 1.

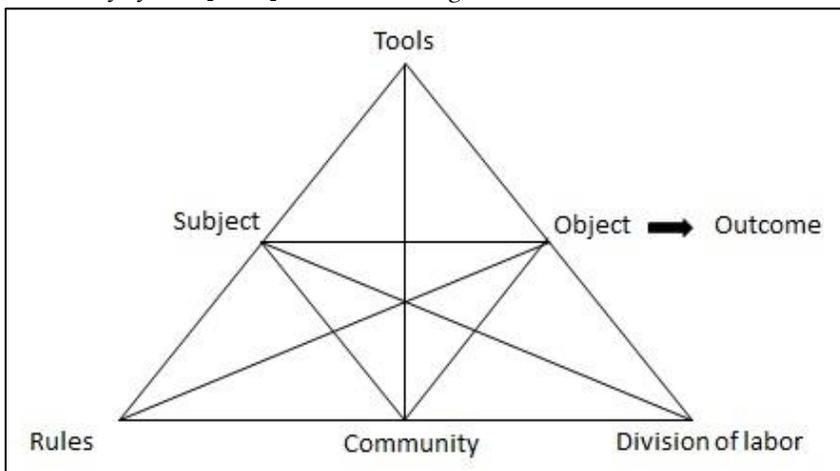

**Figure 1. Activity System** (adapted from [25]).

The activity system can be used to analyze how an activity is culturally and historically situated in a community and collectively shaped by community history and shared understandings. An activity system is a stable one if all the components could maintain a delicate state of equilibrium. However, as people and community change and technologies keep evolving, contradictions, are likely to emerge within an activity system [23,25,50]. Contradictions are opposite tensions that both rely upon and negate one another, such as breakdowns of one element in the activity system, or incompatilibities between two elements. Contradictions are not shortcomings or problems to be solved, but positive forces for change in social systems [12,25]. Specifically, contradictions emerge in the process of knowledge construction through communication and can be generative in discovering and generating new knowledge [12].

An AT lens considers transparency and accountability issues as inherent contradictions between users and their tools. According to AT, if tools are to serve users, users would consciously seek knowledge about the tools' affordances as well as internal workings in order to use the tools properly and achieve desired outcomes. Issues of transparency hinder the user of algorithmic tools because they prevent users from internalizing knowledge about how tools work. Accountability issues could ensue if users fail to understand how tools work. User-generated explanations as users attempting to develop new activity systems that could resolve contradictions introduced by systematic vagueness. The underlying assumption is that users have the consciousness, agency, and motivation to devise their own activities in order to restore the balance of their activity systems by resolving contradictions.

Use the introduction of a new AI tool in an online community as an example. Prior to the arrival of the AI tool, the online community already engages in certain activities, such as content creation. That is, the online community has formed and maintained an activity system involving community members utilizing an online platform for content creation. Suppose that the new AI tool is designed to automatically recommend similar content by other users to inspire content creation. In this way, the AI tool is a new element in the activity system of content creation, and subsequently engenders contradictions. The contradictions could be people's needs to adapt to the AI tool, or their needs for explanations of what the AI tool can or cannot do in order to make the most out of the tool. These are contradictions that will push systems elements and practices to transform until the contradictions are resolved, or people's needs are satisfied.

As people seek explanations to resolve the contradiction, the activity of explaining AI emerges in the community. In the activity of explaining AI, the subject is an explainer, which could be various stakeholders such as users and developers; the tools are AI system, and information and communication tools that the explainer uses to communicate explanation; the object is explaining AI to explainee; the outcome is the explainee understands how AI works; the rules are the mutual agreement about how explanation ought to be carried out; the community is comprised of people who use the particular AI system on a platform; and division of labor means that different stakeholders such as explainers, explainees, developers, and average users assume different responsibilities.

## 4 BACKGROUND: MODERATION AND AI IN THE COMMUNITY CONTEXT

The community under study is *League of Legends* (LoL), a multiplayer online game developed by Riot Games (Riot for short in later text) in 2009, and one of the most popular games in the world [85].

LoL is a match-based online game. In each match, two teams appear in the opposite ends of a map, and only one team wins (Figure 2). LoL embraces a highly competitive eSports culture [55]. Around the eSports culture, the LoL community has gradually developed behavioral rules, which Riot called sportsmanship [11], including general suggestions such as being friendly to and supportive of teammates, as well as harnessing teamwork.

The intense competition of an eSports game is often seen to be related to its notorious reputation for its players' toxic behaviors such as racial slur, harassment, and personal attacks [57,87]. Riot has dedicated much effort in curbing player toxicity for the past decade. They have implemented and maintained an AI-led moderation system since 2014 [56,76]. Therefore, the community and the AI-led moderation system have co-developed in the past few years, as the community adapts to an algorithmic system in charge of monitoring, adjudicating, and sending penalties based on players' behaviors.

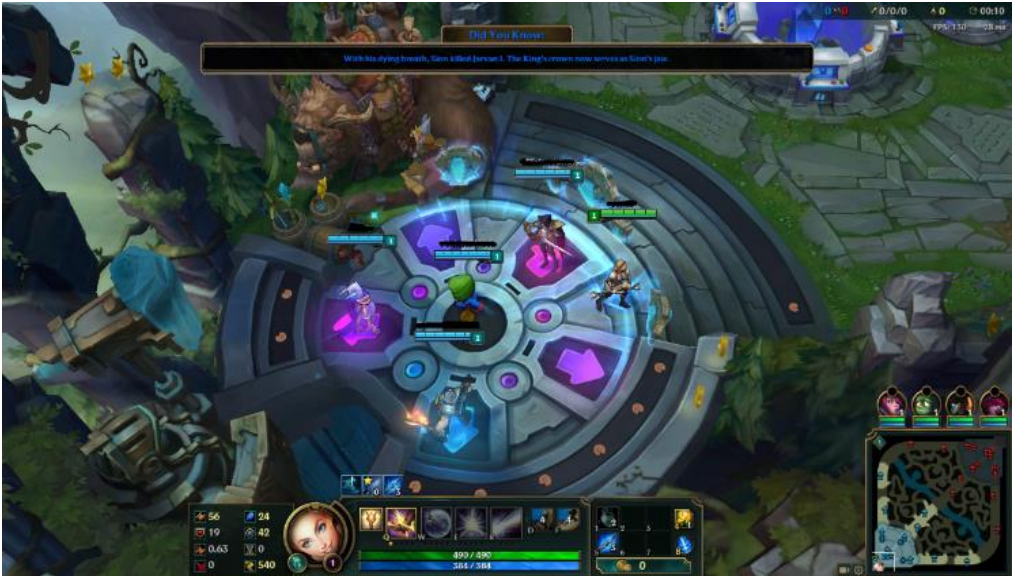

**Figure 2. Screenshot of a 5v5 match in League of Legends.**

The AI-led moderation system's core algorithmic processes operate on their own, and do not have a "human-in-the-loop" mechanism. However, like most current AI systems, the AI-led moderation system does require "labeling" work from the community: Players need to use the "report" function if they perceive toxic behaviors, and thus labeling a particular player's behavior for the moderation system to review. Figure 3 (a) shows the report function, where the player could select from a range of toxic behavior types, as well as add additional comment. This is where players label data (aka player behavior) for the AI system. Once the player clicks the "report" button, this report is sent to the AI system. Figure 3 (b) is an example of a penalty: a pop-up window (or "reform card") for a player who received a penalty, a 14-day account suspension, from the AI-led moderation system. The system determined that the player was toxic, and a suspension was warranted. In addition, the reform card also contained a link titled "Game 1." The link could redirect the player to a webpage including their chatlog from a match, which contained toxic content that could justify the punishment. Figure 3 (c) is a feedback function for players who do the labeling work for the AI moderation system. Whenever a player is punished by the AI system, those who have reported the player will receive a feedback report, acknowledging that they provide input into the AI system, and the AI system has made a decision accordingly.

Only in rare cases would Riot's support team manually review cases and modify penalties, provided that punished players reach out to the support team first. The data the AI-led moderation system uses to make decisions include the reports sent by players as well as the reported players' in-game chats. Riot has not released technical details about the system as to how it makes decisions and determines the severity of penalties.

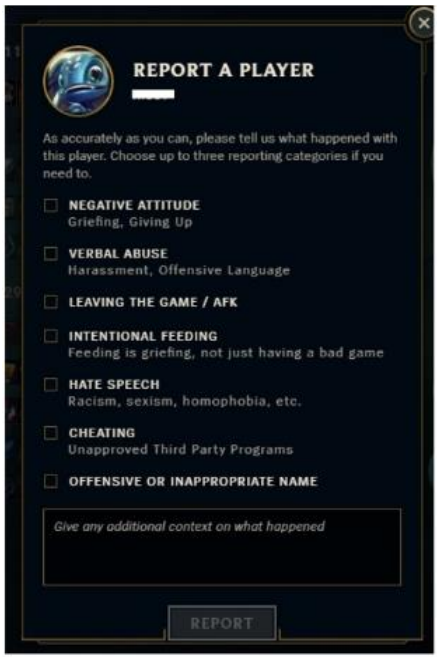

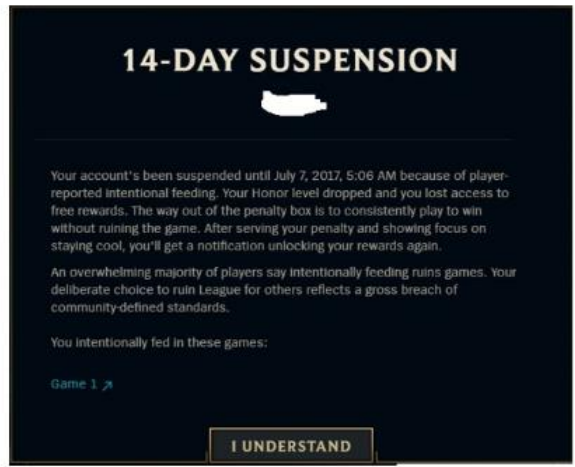

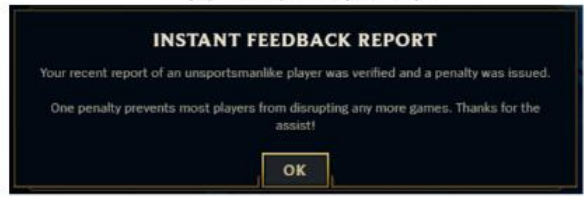

(a) Report function.

(b) Example of penalty.

(c) Feedback to reporter.

**Figure 3. AI-led content moderation system.**

## 5 METHODS

Between January 2019 and February 2019, the researchers who are experienced LOL players collected player discussions from LoL boards (a forum supported by Riot Games' official server). (The boards were discontinued by Riot in March 2020 [33].) We considered player online discussions the most suitable data source for this study for several reasons. First, players who received AI-led decisions (penalties in LoL) were a small portion of the whole player population, about 5% according to Riot's own report [38], and were difficult to identify. There was not an online community, to the researchers' best knowledge, was composed of this type of players. Therefore, it would be improbable to locate these players for an interview or a survey questionnaire. In fact, the researchers had managed to identify punished players (within a larger project studying player experience) for an interview in late 2018. We halted the interview approach after six interviews because we found that in retrospective interviews the six interviewees were unable to fully recall concrete instances of explanation associated with their penalties, most of which happened more than one month prior to the interviews. Instead several of them mentioned browsing LoL boards for answers. Understandably, the semi-structured interview allows participants to recall major experiences or events, but could prove limited in obtaining accurate details of their cognitive processes of constructing an explanation. Thus, we opted to not use the interview approach, and focus on online data. Second, receiving penalties for being deemed as toxic might be considered as a form of social stigma in the community. This would also add challenges to directly soliciting answers from players through interview or survey. Third, existing player online discussions could well support exploratory research efforts because existing research has not yet reported on how people develop explanations for AI, and our bottom-up approach relies on an inductive analysis of data to answer our research question. Fourth, player online discussions are conversations where players collectively explored explanations for AI, resonating with Hilton's conversational model of explanation [43].

We used the forum's search function to locate relevant threads. Based on our understanding of the content moderation system in LoL, we generated an initial set of search terms, including "artificial intelligence," "AI." "suspension," "automated system," "explain," and "explanation." Following an iterative process of reading collected content and refining the set of search terms, we included several more terms such as "report system," "punishment system," "account suspension," "warning," and "ban." We kept collecting data until we reached theoretical saturation, meaning that we could no longer find new concepts and relationships [36]. Our final dataset included 112 discussion threads, with their associated 3970 comments. The most commented thread had 768 comments, while the least comment thread one.

We used qualitative analysis techniques including open coding, theoretical coding, and selective coding [36] to analyze our dataset. The open coding processes involved two researchers assigning initial codes to all the ideas expressed in the dataset. The unit of analysis was either a sentence or a paragraph, depending on how succinct or loose a player expressed their idea. Then they used theoretical coding to building higher level categories and concepts, through a combination of inductive and deductive thinking. Lastly, they conducted selective coding to focus on the core categories reflecting the focus of this study, which is the nature of explanation players developed for the AI system. The final concept map includes three overarching types of explanations: socially-oriented, system-oriented, and action-oriented explanations. Each explanation category contains secondary categories. The AT was not used to directly inform the development of codes as it serves as a grand, high-level theoretical framework. Our later interpretations in findings, however, involve the AT lens.

We consider ethical implications of using online data in this research. The university IRB approval was obtained prior to this study. We believe that our use of this online data introduces little to minimal risk to LoL players because their online utterances do not contain sensitive or personally identifiable information. We do not anticipate that any possible forms of harm that would ensue. Meanwhile, the data could help provide unique insights into the explanation of AI systems. When reporting data, we used rewording to both retain the original meaning of the quotes, while reduce their searchability.

## 6 FINDINGS

We found three types of explanations through which players construct knowledge about their activity system. First, socially-oriented explanation supports players to communicate about rules in their community. Second, players explore system-oriented explanations to develop knowledge about how AI worked. Third, players use action-oriented explanation to resolve contradictions between them and their received penalties to develop new activity systems.

Recall that an AT lens considers the pursuit of explanation as an activity carried out by both explainers and explainees. The three explanation types specify different contextual information that people drew from to construct an explanation.

### 6.1 Developing Socially-Oriented Explanations

The socially-oriented explanation concerns the elaboration on rules and community in the activity system. The pursuit of this type of explanation allows explainers to clarify punishment rationales in light of social values and norms.

After Riot implemented AI to enforce the general behavior rules, AI has become a new tool within the existing activity system of playing LoL. That is, when players play the game, they are aware of and impacted by the AI system. The AI moderation system has brought contradictions into the LoL community by contrasting a few players' beliefs or understandings of their behavior with general behavior rules within the community. Some players question the decisions of the AI system, while others seek to resolve the contradictions by aligning rules with AI-led decisions. An example is:

*Player1: I have suspended account on 14 days. I cannot play on this account any more. What happened to my account?*

*Player2: You broke some rules since you received a punishment. Either hate speech/usage of racial/homophobic slurs or trolling usually results in a 14-day ban.*

In this example, Player1 sought explanations for the AI-led penalty. Player2 explained that Player1 must have broken rules. Player2 reasoned in this way based on a causal relationship between breaking some behavioral rules and receiving a punishment, suggesting their trust in AI decision. Player2 then went on to indicate certain types of toxic languages which Player1 might have used. Behavioral boundaries were clearly drawn that hate speech, racial/homophobic slurs, and trolling were considered toxic. If Player1's question pointed to the contradiction about how the AI system disrupted their routine gaming experience, then Player2's explanation was to resolve the contradiction by aligning AI's rationale with the general behavioral rules held by the LoL community.

Punished players sometimes specify the languages they have used in game when seeking advice. Here is an example:

*Player3: The one time that I retaliate, in a relatively mild toxic manner, I get a 14-day suspension and it is only evident in the chat log of one game! I called a Wukong a braindead monkey for splitting when the enemy team was sieging mid and it was at the very last second of the game.*

*Player4: Braindead monkey is not a "relatively mild toxic manner".*

Wukong is an LoL character. Player3 and Player4 explored the toxic degree of a very specific term, "braindead monkey." Clearly, Player3 and Player4 had different interpretations. By stressing the highly toxic nature of this term, Player4 expressed agreement with the AI decision of 14-day account suspension.

Lastly, players may post their reform cards containing their in-game chatlogs and seek explanations from the community. The community then sifts those chatlogs to understand why the AI system deemed the player toxic. An example is:

*Player5: Does the suspension seem fair, as a first-time punishment, given the chat log (given me being argumentative/toxic during the game)?*

*Player6: Yes, it is fair. There is never any reason to repeat what someone else says. By repeating their hate speech, you become their mouthpiece... Take it as a lesson to not try and take justice into your own hands because you are not qualified to do so.*

In this example, Player5 shared their reform card for advice. In the chatlog, Player5 retaliated using similar hate speeches. Player6 supported the AI decision, and told Player5 that using hate speech under any circumstance would be toxic.

## 6.2 Seeking System-Oriented Explanations

Unlike the socially-oriented explanation, the system-oriented explanation does not draw primarily from behavioral rules or community values. It seeks behind-the-scene knowledge about the AI system, such as how the developers think and design. It originates from the division of labor where the platform owner maintains full control of such knowledge. But the community learns, and shared knowledge of the AI system accrues. The shared knowledge is generated and transmitted through the use of tools, such as the forums that punished players used to seek help, the interactions they had with the AI system, as well as the specific language they used to seek explanations.

The AI-led moderation system is opaque, leading to contradictions between people's desire to know and the tool's opacity. People's explaining activity seek to resolve such contradiction by explicating the rationales of AI decisions. System-oriented explanations refer to all the design characteristics of the AI system that could help make plausible explanations, including design philosophy of the AI system, algorithmic mechanisms, and algorithmic parameters.

### 6.2.1 Design philosophy of the AI system

Design philosophy refers to the overarching goals intended to achieve as well as the primary values intended to uphold in the system. The design philosophy aligns with Riot Games' interests and policies about player behavior and does not concern implementation details of the AI system. LoL players frequently consult information at this design philosophy level to explain AI. For example, two players explored the legitimacy of account suspension:

> *Player7: Hi Riot, I live in America which is a free Country. So, fix your games suspension and banning rules, or take your game out of America. I am sick of getting unnecessary bans, and they are not even reviewed by Riot, but by other players which is absolutely absurd. Fix this or you will lose all of your player base and have many future problems. Thank You*
>
> *Player8: If you think your suspension was unnecessary, please feel free to post your chat logs and prove it. Private companies have every right to control their customer base - including removing them from the "premises" if they so choose. Bans are not reviewed by players; bans are issued by Riot through the automated Instant Feedback System. If you want a human review, you are free to contact support...*

In this case, the explanation that Player7 sought for their suspension stayed at the ideological level of whether a private company like Riot Games has the right to censor users' speech. The player was not seeking specific explanations about whether their behavior was toxic or the system factors that led to their account suspension. Therefore, answering such question required a look into the design philosophy of the AI system.

In some cases, players refer to Riot Games' player behavior policies when explaining system decisions, with the assumption that such policies have been incorporated into the system's decision making. Riot's policies can be viewed as rules in an activity system. Below is a relevant conversation:

> *Player9: I was just placed on a 14-Day suspension because teammates flamed me and called me a dirty %%%%%... They reported me and I received a 14-day suspension when I didn't say* *anything rude to them. Is there a way I can appeal this because the decision was reached in error?*

*Player10: There is no possible scenario in which this would penalize you. You cannot be penalized for what other people say/do. You are responsible for your own words and actions. Riot Games has a zero-tolerance policy towards hate speech. Whoever types it is responsible.*

In this case, Player10 explained that at the design philosophy level, the AI system was enforcing Riot Games' zero-tolerance policy. This means that hate speech led to a 14-day account suspension. However, Player10 did not need to go into details about how such policy is implemented.

Players could also align their own values with how Riot Games designed the AI system. Here is an example:

*Player11: Why is riot so bitchy about shit talk? League is literally the only game I ever have played that bans people for talking shit.*

*Player12: Slurs are not shit talk. Slurs are, by definition, offensive language. It isn't wrong of a company to say "freedom of speech" and let people say anything, imo. It ALSO isn't wrong to enforce rules in an attempt to make the community a better space for everyone (except privileged white dudes, which I am guessing you are). Don't get me wrong, I am also a privileged, hetero, cis, white male. The difference between you and I is that I know it, and I attempt to keep the world nice for everyone, not just white males.*

In answering Player11's question about why the AI system functioned in a way they considered rigid, Player12 shared his own behavioral principles and understandings, and interpreted what Riot Games hoped to achieve through the AI system. For him, the design philosophy of the AI system was to make sure that all the players could "treat each other with respect."

### 6.2.2 Algorithmic Mechanisms

Algorithmic mechanisms are concrete design guidelines or principles that players believed the AI system followed in decision making. Algorithmic mechanism is more concrete and less abstract than design philosophy, often concerning the goal and principle of one algorithm or a set of algorithms. If design philosophy explains to players why the AI system exists and functions in a particular way, then algorithmic mechanism is about the concrete causal factors the AI system uses to convict a player. Here is an example:

*Player13: the punishment system relies on scanning for code words and ignores context riot is saying its ok to report people for saying "fuck" even if it's "fuck yeah!"*

*Player14: That will reduce your report weight to nothing, and the system will ignore your reports.*

*Player15: That's not what context means. I think that's what you're misunderstanding... Context means either the words surrounding the statement, or the circumstances in which the statement was said. If taken out of context, "kill yourself" is a negative phrase. In the context of someone saying "if you don't get some lifesteal you will kill yourself on his Thornmail", it is not negative at all. If you ignore the context, you can't tell whether it's bad or not. If you pay attention to the context, you can tell. This program pays attention to the context.*

"Lifesteal" is a property of in-game items that allows characters to gain health points while attacking the enemies. "Thornmail" was a type of in-game item that reflects damage onto the attacker. In this conversation, the three players' discussion revolved around two algorithmic mechanisms: report weight and context. Player13 hypothesized a scenario of "abusing" the report button without considering the context of certain behavior. For example, some languages might be toxic in the community, but acceptable in a smaller group of friends. Player13 questioned the sophisticatedness of the AI system, asserting that the AI system was purely rule-based and ignored the context of player behavior. Player14 explained by following Player13's logic and drawing from their understanding of algorithmic mechanism, suggesting that Player13 has a misconception of the AI mechanisms, and that Player13's proposed action would dynamically alter how the AI system evaluated Player13's subsequent reports. Player15, on the other hand, argued against P13's statement about the AI system ignoring context. Player15 explained in detail how AI incorporated context when making punishment decisions.

Player discussions also touched upon the learning capacity of the AI system:

*Player16: do you know how fucked up that is? This game is known for its toxic community and you can see why.*

*Player17: It's not the words that the system detects, it's the context too. Just because one person is raised somewhere where the language filter and racial slur filter is more lax it doesn't mean the rest of the server should have to listen to it. The report system is literally adjusted based on what WE say is unacceptable. If all of us stopped reporting "gg ez" or "ur scrubs" then it would probably fly right under the system's radar completely. WE made the list; Riot just compiled it.*

"gg ez" is short for "good game easy," which is used by winners and often deemed as offensive to the losing team. In this conversation, Player17 acknowledged that regional standards for the toxic/acceptable boundary could differ but stated that the game community had universal standards for player behavior. To Player17, it was the player community that fed raw data into the AI system, which then learned behavioral norms. The algorithmic mechanism in this discussion is how the report system worked hand in hand with the punishment system.

Although one account suspension appears as a one-time punishment, players perceive it as merely the tip of the iceberg. They collectively discuss how the AI systems has made many other decisions that remain hidden from them. For example, several players discussed:

*Player18: Does anybody know if the suspensions disappear after a certain amount of time, or do they remain permanent on your record?*

*Player19: it sounds like it takes around 3 months to drop a punishment tier. So, you are on tier 3 in around 9 months you will be back to a clean slate. if you don't get punished again.*

*Player20: This is dangerously untrue. It's both a combination of time (and there is no set time) and games played without negative behavior (again, no set number).*

According to Player19 and Player20, once a player received a punishment, the player no longer had the "innocent" status. Instead, some watchlist mechanisms were in play, closely monitoring them. The mechanisms could involve a diversity of factors.

### 6.2.3 Algorithmic Parameters

Algorithmic parameters are peculiar thresholds that the AI system uses for boundary detection. Algorithmic parameters are related to mechanisms but are more specific and oftentimes focused on variables such as numbers and dates. Here is an example:

*Player21: My punishment says it ends today August 17th and my account still suspended... Hopefully you guys can see this and give me a straight answer.*

*Player22: Bans end at the exact time they were issued. The time the email was sent to you should act as a good basepoint for that. Remember, though, that Riot is on PDT, so you may need to convert it to your local time.*

Here Player21 sought explanations for why their punishment had not ended. Player22 referred to the specific timing parameter (time zone = PDT) which could help explain Player21's punishment status.

Players refer to algorithmic parameters to explain concrete questions about punishment severity. Here is an example:

> *Player23: I haven't gotten a ban in about a month and a half, then i get mad 1 game. 1 GAME in almost 2 MONTHS... and now I'm 14 day banned. Please riot show some mercy for those who dont deserve it.*
>
> *Player24: Takes more than one game to get banned. Especially since bans require two previous restrictions, therefore the theoretical minimum is THREE games if one report COULD get you in trouble.*

In this example, Player24 disagreed with Player23 upon what frequency of behavioral violations led to a punishment. Player 24 proceeded to detail several numbers that would be used to determine toxicity and punishment level. It is unable for outside researchers like us to verify whether these numbers reflect how the AI system actually works. However, we could observe from this example that players used algorithmic parameters as a rhetorical tool to help their argumentation.

Lastly, players would also assign parameters to patterns they believed they had observed on the AI system. Here is an example:

> *Player25: every player will get a message normally within 15 mins if the player did get punished for those that did report. As I have, and several teammates have got the message shortly after we reported a player as we got into another Que. That doesn't mean the player did get banned but they could have got a chat restriction, warning, or a longer que time. The punishment could also take about 15 mins to take in effect as well.*

In this example, Player25 confidently stated 15 minutes as a temporal pattern for the system to make decisions. Player25 made this statement based on their previous interactions with the AI system.

## 6.3 Identifying Action-Oriented Explanations

The action-oriented explanation reflects the activity system's inner quality in producing new activity systems. As people receive and reflect upon system results, they can exert agency and creativity to devise new activities for better system results.

In LoL punished players do not enjoy penalties, and desire to keep playing the game. The contradictions between their desire and penalties motivate them to seek explanations that can help them avoid penalties in the future. In doing so, they create new activity systems such as repairing and reforming. As such, the explanations generated are forward-looking and actionable.

### 6.3.1 Repairing

Players seek to repair if they believe their punishments result from an error in the AI system. The repairing explanation usually uses probability language to suggest what possible actions could be done to remove or modify a penalty. The only way to repair is to contact Riot Games' player support team. Here is an example:

> *Player26: But the permanent suspension is so severe given I just tested the cheating script out of curiosity in practice mode. Can I somehow get a temporary suspension or ban?*

*Player27: ... I can only recommend submitting a Support Ticket, but I won't try to claim that you have a good chance of successfully appealing the punishment.*

In this conversation, Player27 pointed to repairing as a possible future action for Player26. To resolve the contradiction of a perceived unfair punishment, P26 needed to initiate a new activity of contacting Riot's support team.

In another example, players further articulated when repairing would help deal with the punishment:

*Player28: Could my suspension be removed?*

*Player29: It depends. Punishments are generally not removed/lifted unless there was some form of error on Riot's end in giving them... sometimes people will get a punishment when they truly don't deserve it. It's very rare, but it can happen, and when it does Riot is more than willing to reverse the punishment.*

In this example, Player29 explained that the AI system was rigid about its punishments, and that Riot Games had the power to override the AI system. However, repairing was only possible when Riot Games deemed a punishment unreasonable. For example, "kill yourself" was generally considered offensive. However, as Player15 put it, the phrase could be inoffensive when the speaker was referring to a specific in-game context. If the AI system labeled "kill yourself" as an offense but missed the particular context, repairing was a possible action.

### 6.3.2 Reforming

Players seek to reform their behaviors if the AI decisions have been justified. When players' conversations reach an agreement on an explanation of why the AI system is correct and a penalty is justified, a follow-up discussion sometimes ensues about what the player should do in the future. The reforming explanation reasons about what actions could be taken in the future to avoid future punishments. Below is a brief conversation on this topic:

*Player30: Is there any way I can have my suspension removed. I apologize for my bad behavior and will work to never chat with anyone ever again in league of legends since 90% of the time it is negative.*

*Player31: You must be new here. They're not going to remove the ban. But take this as a valuable lesson because the next punishment, even for mild toxicity, is going to be a permanent ban. Let's not see you back here posting in 2 weeks.*

In this example, Player31 suggested that Player30 learn from the present punishment, and improve their in-game behavior. In another example, Player32 provided concrete suggestions for how to reform and avoid further punishments:

*Player32: Any account can go back to have a clean slate in time. at least what my friend told me, he's been playing a while, so I trust what he says. But that clean slate may take several months. (Especially if your reforming from a 14-day ban.) As far as End of season stuff goes, you still qualify with just a merely chat restriction. But be sure you won't get banned anytime soon. <3 My advice to you? Keep /all chat off. I turned mine off and it's been a more enjoyable experience. Sometimes it's also best to just not talk at all. You keep focus and end up doing better...*

"/all" is the option in LoL that allows players to talk to the opponent team. If this option is turned off, players will no longer see messages from the opponent team. Player32 suggested turning this option off, a concrete in-game act, as a strategy for avoiding provocations that could lead to toxic behaviors.

# 7 DISCUSSION

In this study, we analyzed explainable AI from an activity theory lens, and used a case study to illustrate how the activity of explanation played a role in human-AI interaction, and more specifically how situated explanations draw from community values and rules to mediate community-AI interaction. Explanation is not just about providing accurate information about how AI works, but fundamentally social and situated. The community-supplied explanations help resolve contradictions for individual players who seek information as well as other community members who seek to deepen their shared understandings of the AI system. The three explanation types, namely socially-oriented, system-oriented, and action-oriented, are not meant to be mutually exclusive. For instance, a system-oriented explanation may carry socially-oriented values, like the conversation between Player11 and Player12.

## 7.1 The Role of Community in Explainable AI

In our social analysis of XAI, the emphasis is not on whether certain explanations are persuasive and effective, but on the social processes through which explanations are generated. For social theories like AT, the "independent variables" at the center of analysis are those social processes, while the effects of those processes are considered "dependent variables" that are highly contextual and subject to change. When the community provided an explanation to clarify a player's penalty, the composition and logic of the explanation depended very much on the player's specific behavior, how the player framed the question, as well as who were available at the moment to provide answers. Such explanation may or may not appear sensible to the player. But most importantly, the LoL boards, the place where explanation took place, facilitated such a social process where players could collectively figure out explanations, and our analysis could identify three core explanation types afforded by the social process. Use the conversation between Player11 and Player12 as an example. The emphasis of our AT analysis is not how many players in the community are like Player11 or Player12, but that the LoL boards provided a learning opportunity for players like Player11 to reflect upon and improve their behavior.

The AT analysis of explanation traces explanation's origins in the social, cultural, and historical conditions [25,50]. The activity system of explanation in the online community entails the social construction of knowledge through communicative practices as documented in our findings. Therefore, understanding explanation as a social process means first to acknowledge the roles and agency of various social actors with different interest and viewpoints, including multiple explainers and explainees and platform owners. In other words, the development of sound explanations and XAI should involve voices from multiple stakeholders.

Second, explanation is conditioned in existing sociotechnical arrangements. For example, punished players utilized the existing player behavior forum to discuss explanations. Thus, explanations of existing AI systems are likely to have already taken place among their users, and AI developers could and should turn to their user community for understanding the role of explanation.

Third, discourse and power are at play in the explanation process. For example, players acknowledge Riot's authority in explaining the AI system, and their collaborative explanations often revolved around Riot's platform policies. The AT lens suggests that explanation happens in and draws insights from a particular community or culture. For example, what comprises toxic behavior involves rules pertaining to a particular community, and varies across different communities. In LoL, "kill yourself" could be acceptable in a particular in-game communicative context where teammates warn players that if they make bad decisions, they may get their characters killed in game. This phrase could appear highly offensive in other cultures.

Fourth, the AT lens also considers the historical conditions of explanation. Explanation is not an isolated event in the temporal continuity. Players compared the human-led and AI-led moderation systems [54]. They developed explanation from historical events such as their previous experiences with the AI system and explanations. Their explanations were also informative for future evolvement of activity systems by players who were more disciplined in gameplay.

This study has placed community at the center of analysis, pertaining to the basic assumptions of AT. In so doing, we have associated the collectively generated explanations with certain shared values and beliefs upheld by the LoL community. However, researchers from CSCW have also critically reflected upon the notion of community, acknowledging that norms in a particular online community could have peculiar norms that deviate from societal values [2], and that even successful online communities like Wikipedia might have already developed values and norms that marginalize newcomers and minority groups [10,41]. Elsewhere, scholars have criticized the gamer culture for being masculine and toxic [67], and LoL is notorious for its players' toxicity [61]. Therefore, the notion of community can be critically approached, and this is partially why community moderation has become an important topic in recent CSCW literature (e.g., [6,13,14]). That is, community norms are not static but dynamic and evolving [90], and moderation could steer community norms towards positive changes by impacting the shared understanding of what behaviors are acceptable or unacceptable [58]. Players who have committed toxicity deserve an explanation to understand why their behaviors are toxic and have the chance to improve their behavior in future gameplay. Thus, we consider explanation a critical social process in educating toxic players, and subsequently transforming problematic community norms.

## 7.2 Reconciling Contradictions through Explanation

A salient theme across our findings is the emergence of contradictions due to the introduction of AI moderation into the community. The introduction of new element is a common cause for contradictions [12,25]. Riot Games initially envisioned the AI system as a tool that could help discipline toxic players (toxicity itself is another contradiction). Such new element has destabilized existing activity systems which players were accustomed to, and contradictions arose. Such contradictions were primarily epistemic ones focused on ways of knowing rules by players and AI. Our findings demonstrated how players sought to use explanation to reconcile the contradictions within their activity systems.

Recall that transparency and accountability present inherent contradictions in complex systems. JHaver et al. showed that on Reddit, content removal explanations provided by authorities such as the platform and human moderators could help users learn about social norms and reduce the odds of future post removals [47]. Our findings similarly pointed to opportunities for learning social norms in the socially-oriented explanations. Additionally, our inductive approach revealed the complexities and nuances in user-driven explanations, and the relation of the community context to situated explanation. For instance, punished users also desire system-oriented explanations about moderation systems' decision making processes, as well as action-oriented explanations concerning their selves.

An ideal human-AI interaction scenario would be a stable activity system, in which humans smoothly interact with AI systems for their own activity goals. However, the inherent vagueness and complexity of AI [3,19,80] naturally leads to epistemic contradictions which could only be resolved through explanation.

By analyzing explanation as activity and considering explanation's interactive nature [43,69], we stress how contradictions entail the various forms of explanation. Because contradictions involve oppositional forces and tensions, explanation often involves multiple entities and back-and-forth exchanges of ideas and information between them. Explainer, for example, could be comprised of one user, multiple users, the platform, or any combination of them. Explainee could be one or multiple users. The conversational patterns could be either unidirectional or bidirectional. Explanation could happen pre-hoc or post-hoc. For example, Riot Games performed a pre-hoc unidirectional explanation by announcing how the AI system enforced rules. Players who received punishments performed post-hoc explanation within the community but also drew from Riot Games' explanation.

## 7.3 Rethinking Explainable AI through Activity Theory

Both technical efforts of building explainable AI models [39,70] and cognitive science informed framework [99] of how human minds process explanation tasks seek to provide expert-driven explanations. The assumption is that users will understand AI systems so long as the information with the right quality and quantity is provided. In this paper, we analyzed XAI from an AT lens, foregrounding explanation as human activity and stressing the contextualization of explanation.

We understand expert-driven explanatory efforts as tools that are inserted into users' existing activity systems. Tools could play a mediational role if tools do not require additional understanding and learning from users [93]. Tool embodiment describes the tight, unconscious coordination between the body and the tool, such as how people use hammers [95]. However, tools such as technical XAI systems may cause new contradictions. We use Figure 4 to map out how a purely technical XAI approach might interact with an activity system. If a technical XAI system were deployed in the LoL community to explain the workings of the automated moderation system, the XAI system would not be smoothly incorporated in players' existing activity system of explanation. There would be multiple contradictions: one between subjects and tools, as users do not necessarily understand technical XAI (i.e., a simpler AI model [84]); one between users and division of labor, as users have agency and will develop their own reasonings; one within tools, as users might seek alternative tools, and find contradicting explanations; one between tools and rules, as XAI tools might not conform to existing rules; and etc. Although AT theorists see contradictions as positive forces for change [12,25], this however might seem as an endless loop where new tools lead to new, nested activity systems of explanation.

Therefore, the AT lens enables us to ask how XAI could strike an equilibrium in an activity system of explanation by considering the six components and their inter-relationships. First, XAI should consider how users and platforms could work together as *subjects*. As such, the *division of labor* should also create situations to allow users to explain and platforms to listen. This entails collaboration and common ground. Platforms should not act merely as authorities. If users and platforms work as a team, then technical XAI systems and developer-driven explanations will no longer be distant from users.

Second, XAI should consider the reconciliation of the diverse explaining *tools*. It is no longer enough to devise a single communication channel when users themselves are utilizing multiple means to develop explanations. In our findings, users are utilizing networked resources including, but not limited to, their own experiences with the moderation system, their interactions with the game client, their observations of others' behaviors, their understandings of community rules as well as Riot's interpretations of them, and Riot's explanations of the moderation system to develop explanations. Within these means there are rich opportunities for Riot to communicate with players regarding how the moderation system functions, beyond the online forums that players are already using. For example, Riot could host open discussions about the system, formalize communication channels where players could directly consult the designers, or support dedicated social venues for players to share experiences and help each other improve.

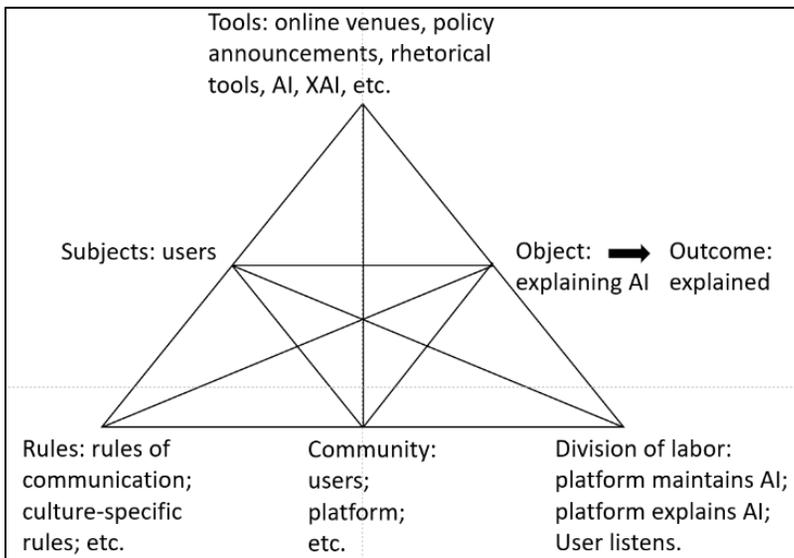

**Figure 4. An activity system of XAI.**

Third, the *object* is not just explaining AI, but also situating and contextualizing AI, making explicit the circumstances such as rules and community that co-determine an AI decision, and incorporating an AI system into the community life. When community managers consider introducing an AI system into an existing community, explanations happen dynamically and continuously through the lifecycle of the community. There is unlikely to be a one-time magic trick that could immediately resolve all the questions and doubts from the community.

The contextualization of explanation looks beyond the spatio-temporal bounds of a single explanation. In our study, players actively built links between the penalties at discussion and their previous experiences and other competing explanations. This suggests that explanation is not isolated but connected. In terms of AT, the activity systems of explanation in a particular community are overlapping and nested. The implication is that an XAI project should consider previous explanations as well as concurrent explanations while preparing a current explanation.

## 7.4 Implications for Design

Design of human-AI interaction needs to consider how to incorporate explanation to maintain an equilibrium in the activity system of community-AI interaction. Our former AT analysis of XAI synergizes well with the notion of participatory design (PD) [74], in emphasizing users' involvement in the design. Echoing with existing discussions at the intersection of PD and algorithmic design [21,60], we suggest that various types of stakeholders should participate with platforms in the design process. More specifically in the context of AI-led moderation, both punished players and "innocent" players should be invited to work with the platform to explore quality explanations. Second, participation should occur iteratively throughout the development life cycle, so that users and platform could collectively identify and resolve contradictions at their early stages. Ethnographic methods and inviting users to envision possible explanations could be deployed to understand the context where AI will be deployed. At a later stage, users should work with platform together through processes of prototyping and evaluation. Our study highlights how empirical investigations could reveal situated user perspectives in the wild.

The activity checklist provided by Kaptelinin and Nardi [50] is also informative here. When designing explanation into community-AI interaction, we could ask provocative and generative questions from four perspectives: First, to what extent explanation tools facilitate and constrain users' goals and how they impact users' interactions with AI? Second, how could explanation tools be integrated into users' environments composed of other people, resources, and technologies? Third, could explanation tools facilitate their mutual transformations with users where users could internalize deeper understandings of AI and help improve the tools? Fourth, how would explanation tools develop into future use?

More specifically, moderation systems could consider how to design for user-driven explanation. Informed by ideas such as deliberation and procedural justice, Fan and Zhang implemented a stage-based model for adjudicating content moderation cases [29]. In a similar vein, moderation design could consider beyond the typical content moderation cycle, what possible explanation procedures could be designed to allow community members to collectively develop explanations and learn from each other.

## 8 CONCLUSION

In this paper, we drew from activity theory, a foundational HCI theory, to analyze the mediational role of explanation in community-AI interaction. We presented an empirical case to further illustrate the AT analysis of explainable AI. We argue that explanation is not just about providing accurate information about the internal workings of AI, but also related to the community context where norms, practices, values, and knowledge are cultivated along the use of AI. Explanation is human activity that is socially, culturally, and historically situated. An AT lens that pays attention to contradictions and balanced states is invaluable in understanding and designing sustainable social systems of community-AI interaction.

### ACKNOWLEDGMENTS

We thank the anonymous reviewers for their cogent and constructive feedback.